\documentclass[aps,pre,reprint,superscriptaddress,longbibliography]{revtex4-1}
\usepackage{scrextend} %,wasysym}
\usepackage{amsmath,amssymb,graphicx,subfigure,color,times,tabularx,hyperref,fancyhdr}
\usepackage{esint}
\begin{document}
\title{Analytic Boundary Terms for Arbitrary Crystal Geometries and Direct-Sum Evaluation of Madelung Constants in Triclinic Lattices}
\author{Yang He}
\affiliation{Key Laboratory of Laser \& Infrared System of Ministry of Education, Shandong University, Qingdao 266237, P. R. China}
\affiliation{Qingdao Institute for Theoretical and Computational Sciences (QiTCS), Center for Optics Research and Engineering, Shandong University, Qingdao 266237, P. R. China}
\author{Zhonghan Hu} \email{zhonghanhu@sdu.edu.cn}
\affiliation{Key Laboratory of Laser \& Infrared System of Ministry of Education, Shandong University, Qingdao 266237, P. R. China}
\affiliation{Qingdao Institute for Theoretical and Computational Sciences (QiTCS), Center for Optics Research and Engineering, Shandong University, Qingdao 266237, P. R. China}
%\date{\today}
\begin{abstract}
The direct-sum evaluation of Madelung constants is complicated by the conditional convergence of lattice sums, which gives rise to a shape-dependent boundary term.
In this work, we present, for the first time, a closed-form analytic expression for this boundary term that is valid for arbitrary crystal geometries.
For general triclinic lattices, this boundary term maps exactly onto the electrostatic potential generated by a set of uniformly charged parallelograms.
In addition, we demonstrate that the residual finite-size correction for a crystal of characteristic size $p$ decays as $(2p+1)^{-2}$.
Building on these results, we develop a robust direct-sum method for the accurate computation of Madelung constants in arbitrary triclinic lattices and validate its effectiveness through explicit calculations on representative Bravais lattices.
\end{abstract} \maketitle

%%%%%%%%%%%%%%%%%%%%%%%%%%%%%%%%%%%%%%%%%%%%%%%%%%%%%%%%%%%%%%%%%%%%%%%%%%%%%%%%%%%%%%%%%%%%%%%%%%%%%%%%%%%%%%%%%%%%%%%%%%%%%%%%%%%%%%%%%%%%%%%%%%%%%%%%%%%%%%%% Introduction
\section{Introduction}
Simulations of condensed matter systems frequently rely on periodic boundary conditions (PBCs), wherein a unit cell containing charges is embedded within an infinite array of periodic replicas.
This approach is foundational across molecular dynamics\cite{Allen_Tildesley2017,Frenkel_Smit2023}, electronic band structure calculations\cite{Singleton2001}, and quantum Monte Carlo simulations\cite{Cazorla_Boronat2017}. 
Under PBCs, the central computational challenge is evaluating the electrostatic potential at a reference charge arising from the infinite lattice. 
The computation of this Coulomb lattice sum, historically known as the Madelung series, is a problem dating back over a century\cite{Madelung1918,Ewald1921,Born_Huang1954}. 

A fundamental mathematical subtlety of the Coulomb lattice sum is that the series is only conditionally convergent. 
Consequently, its value depends explicitly on the order of summation, which physically corresponds to the shape of the crystal. 
This shape dependence was initially overlooked in Ewald's original formulation\cite{Ewald1921}. 
The necessity of a rigorous mathematical treatment was first recognized by Redlack and Grindlay\cite{Redlack_Grindlay1972,Redlack_Grindlay1975}, who demonstrated that the lattice sum must include an additional shape-dependent term beyond the standard periodic Ewald expression.
Subsequent mathematically well-controlled derivations by De Leeuw, Perram, and Smith further solidified this understanding\cite{DeLeeuw_Smith1980,Smith1981}.
However, obtaining an explicit analytical form for this shape-dependent term has proven to be a formidable challenge.
As Smith notably remarked\cite{Smith1981}, ``in spite of some diligent work, no analytic progress has been made with [the shape-dependent term].''
While a few analytical results have emerged in recent years for highly symmetric crystals\cite{Ballenegger2014,Pan2017,Zhao_Hu2026}, a general closed-form solution for arbitrary crystal geometries has remained elusive.

\begin{figure}[!htb]\centerline{\includegraphics[width=7.5cm]{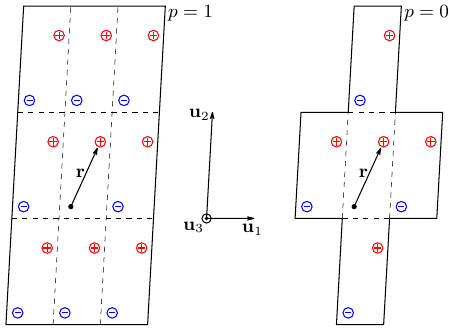} }          %---------------------  Figure -------------------------------------
\caption{Cross sections of two finite crystals in a triclinic lattice (primitive vectors ${\mathbf u}_1$, ${\mathbf u}_2$, and ${\mathbf u}_3$).
Each unit cell contains unit point charges $\pm 1$ separated by ${\mathbf r}$. 
Solid lines mark the exact boundaries; black dots denote the negative ions in the central unit cells. 
Left: A regular crystal of size $p=1$ ($27$ unit cells) bounded by $6$ parallelograms.
Right: An irregular crystal of size $p=0$ ($19$ unit cells) bounded by $30$ parallelograms.
See section~\ref{sec:shape} for the precise definition of exact shape and size.} \label{fig:tshape}\end{figure}
Recently, significant progress has been made by demonstrating that the definition of exact shape and size for a finite crystal enables a clear separation of boundary and finite-size effects\cite{Zhao_Hu2026,He_Hu2026}.
Fig.~\ref{fig:tshape} shows schematic representations of regular and irregular finite crystals within a triclinic lattice defined by primitive vectors ${\mathbf u}_1$, ${\mathbf u}_2$, and ${\mathbf u}_3$.
A general lattice vector is given by ${\mathbf n}=n_1 {\mathbf u}_1 + n_2 {\mathbf u}_2 + n_3 {\mathbf u}_3$, where $n_1$, $n_2$, and $n_3$ are integers, with ${\mathbf n}={\mathbf 0}$ corresponding to the central unit cell.
Let ${\mathcal L}(p|{\mathbf s})$ denote the set of lattice vectors defining a finite crystal of size $p$ and shape ${\mathbf s}$; the condition ${\mathbf n}\in {\mathcal L}(p|{\mathbf s})$ thus enumerates all such vectors contained within this specific crystal.
The definition of the shape and size will be given in the next section.
If each unit cell contains a pair of unit charges separated by a displacement vector ${\mathbf r}$, the electrostatic potential experienced by the target negative ion (charge $-q=-1$) in the central cell (see Fig.~\ref{fig:tshape}) is then given by
\begin{equation} \nu({\mathbf r},p|{\mathbf s}) = \frac{1}{r} + \sum_{ {\mathbf n}\neq {\mathbf 0} }^{{\cal L}(p|{\mathbf s})} \left[ \frac{1}{\left| {\mathbf r} + {\mathbf n} \right | }  - \frac{1}{n} \right] , \label{eq:dir}   \end{equation}
which accounts for the Coulomb interactions with all other charges in the finite crystal.
$\nu({\mathbf r},p|{\mathbf s})$ defines the effective pairwise interaction under the periodic boundary condition\cite{Hu2014ib,Zhao_Hu2025}.
This well-defined finite lattice sum decomposes into three distinct components: a periodic bulk term $\nu_{\rm pbc}$ (the standard Ewald term\cite{Ewald1921} or the so-called Ewald sum with the tinfoil boundary\cite{DeLeeuw_Smith1980,Hu2014ib}), a shape-dependent non-periodic boundary term $\nu_{\rm b}$, 
and a finite-size correction term $\nu_{\rm corr}$ that decays to zero in the limit of large $p$\cite{Zhao_Hu2026,He_Hu2026}:
\begin{equation} \nu({\mathbf r},p|{\mathbf s}) = \nu_{\rm pbc}({\mathbf r}) + \nu_{\rm b}({\mathbf r}|{\mathbf s}) + \nu_{\rm corr}({\mathbf r},p|{\mathbf s}). \label{eq:nusep} \end{equation}

While the definition of exact shape and size enables a clear separation of boundary and finite-size effects\cite{He_Hu2026}, deriving an analytical expression for this shape-dependent boundary term remains mathematically challenging.
Previously, two methods based on the divergence theorem and Fourier transforms were proposed for a finite crystal with a regular rectangular-prism geometry in an orthogonal lattice\cite{Zhao_Hu2025}.
However, these approaches fail to yield closed-form expressions for general triclinic lattices.
In this work, we advance this framework to achieve a complete generalization by deriving a closed-form analytical expression for the boundary term $\nu_{\rm b}$ for general triclinic lattices with arbitrary crystal shapes.
Building upon the parameterization of exact crystal shapes and sizes\cite{He_Hu2026}, we employ a new parameterized integral approach that maps the boundary term directly onto the electrostatic potential of a set of uniformly charged parallelograms.
This new method yields an exact, closed-form solution applicable to both regular and irregular crystal geometries across all triclinic lattices (Fig.~\ref{fig:tshape}).
By resolving this long-standing analytical challenge, our work provides a powerful, universally applicable tool for the direct-sum evaluation of electrostatic interactions in condensed matter systems.

In the remainder of this work, we first outline the method for deriving analytical boundary terms in the next section and present the results in Section~\ref{sec:result}.
Details of the derivations involving complex integrals are provided in the Appendices A and B.
Section~\ref{sec:ray} establishes a connection between the conditional boundary term and the classical work by Rayleigh\cite{Rayleigh1892} on the influence of obstacles on the properties of a medium.
Eq.~\eqref{eq:nusep} provides a precise interpretation of the bulk energy in terms of a well-defined finite crystal. 
In the context of bulk simulations\cite{He_Hu2026}, ongoing efforts to develop fast algorithms for $\nu_{\rm pbc}$ focus on extending particle-mesh techniques\cite{Darden1993,Essmann1995,Liang2026,Gao_Greengard2026} as well as mean-field approaches\cite{Hu2014spmf,Hu2022,Gao_Hu2023,Gao2026}. 
As a validation of our formula, Section~\ref{sec:app} demonstrates one particular application to the computation of the bulk energy.
This demonstration uniquely extracts the contribution of both boundary and finite-size effects.
We first apply it to FCC crystals, which were previously studied using larger unit cells in a cubic lattice.
The bulk energy is found to be consistent between regular and irregular crystal shapes.
We then carry out an additional application to the triclinic wollastonite lattice and compare the results with those of the usual perovskite structure.
A brief summary is given in Section~\ref{sec:con}.
%%%%%%%%%%%%%%%%%%%%%%%%%%%%%%%%%%%%%%%%%%%%%%%%%%%%%%%%%%%%%%%%%%%%%%%%%%%%%%%%%%%%%%%%%%%%%%%%%%%%%%%%%%%%%%%%%%%%%%%%%%%%%%%%%%%%%%%%%%%%%%%%%%%%%%%%%%%%%%%%
\section{Boundary Terms of Exact Shapes}\label{sec:shape}
Prior to deriving the analytical expression for a general triclinic lattice, we first present the definition of exact shape and size.
For an arbitrary lattice, we consider a family of finite, centrosymmetric crystals parameterized by a fixed triplet of non-negative integers ${\mathbf s}=(s_1, s_2, s_3)$ and an index $p \geqslant 0$.
The crystal dimensions (in unit cells) are $(2p+1)(2s_1+1)\times (2p+1)(2s_2+1)\times (2p+1)(2s_3+1)$, with the triplet $(2s_1+1,2s_2+1,2s_3+1)$ constrained to be coprime so as to ensure an irreducible aspect ratio.
As $p$ increases, the sequence uniformly expands all linear dimensions, thereby preserving both the centrosymmetric geometry and the exact aspect ratio.
In this setup, the scalar variable $p$ controls the size, while the vector ${\mathbf s}$ fixes the crystal shape; for an orthorhombic lattice, this regular shape corresponds to rectangular prisms\cite{Zhao_Hu2025,Zhao_Hu2026}.
As a two-dimensional illustration, the left panel of Fig.~\ref{fig:tshape} depicts a cross-section for a finite crystal with $p=1$ and $s_1=s_2=s_3=0$.

To maintain exact geometric scaling on a discrete lattice, irregular or composite shapes must instead be constructed from combinations of regular shapes.
In such cases, the shape parameter ${\mathbf s}$ generalizes to an extended set of integers that uniquely define the composite geometry.
The right panel of Fig.~\ref{fig:tshape} illustrates a cross-section of one such composite crystal with $p=0$, whose exact shape is rigorously defined by the union of two distinct regular shapes: $1\times 1\times1$ plus $1\times 1\times 1$ where the latter set denotes the crystals on the six faces.
Alternatively, this irregular shape can be represented by $3\times 3\times 3$ minus $1\times 1\times 1$ where the latter denotes the crystals on the eight corners.
Evidently, this exact shape and size differs from the usual spherical or ellipsoidal shapes that have been widely employed to approximate large finite crystals\cite{DeLeeuw_Smith1980,Smith1981,Kantorovich_Tupitsyn1999,Smith2008}.

Let $V=\left| {\mathbf u}_1 \cdot \left( {\mathbf u}_2 \times {\mathbf u}_3 \right) \right|$ denote the volume of the unit cell in the triclinic lattice.
The boundary term can be expressed as the interaction energy between a point dipole at the origin and a uniformly polarized continuum filling the crystal volume $\Omega(p|{\mathbf s})$
\begin{equation} \nu_b({\mathbf r}|{\mathbf s}) = \frac{1}{2V} \int_{\Omega(p|{\mathbf s})} d{\mathbf x}\,  \left({\mathbf r}\cdot \nabla_{\mathbf x}\right)^2\frac{1}{\left| {\mathbf x} \right|}. \end{equation}
For this integral to be independent of the scaling parameter $p$, the crystal geometry must be parameterized such that its shape remains strictly self-similar during scaling\cite{He_Hu2026}.
This definition of exact shape and size differs from the usual treatment of approximated shapes such as spherical or ellipsoidal boundaries.
Analytically evaluating this integral involves two key steps. 
First, we apply the divergence theorem to convert the volume integral into a surface integral over the boundary $\partial\Omega(p|{\mathbf s})$
\begin{equation} \nu_b({\mathbf r}|{\mathbf s}) = \frac{1}{2V} \oint_{\partial\Omega(p|{\mathbf s})} {\mathbf r}\cdot d{\mathbf S}\, {\mathbf r}\cdot \nabla_{\mathbf x}\frac{1}{\left| {\mathbf x} \right|}. \end{equation}
For a crystal of any given shape described by ${\mathbf s}$, the boundary surface $\partial\Omega(p|{\mathbf s})$ comprises $N_{\mathbf s}$ parallelogram facets, denoted by ${\mathbf S}_1, {\mathbf S}_2, \dots, {\mathbf S}_{N_{\mathbf s}}$, that collectively enclose the region $\Omega(p|{\mathbf s})$.
For example, $N_{\mathbf s}=6$ and $30$ for the two finite crystals shown in Fig.~\ref{fig:tshape}, respectively.
Additionally, $N_{\mathbf s}$ is necessarily even, provided the crystal is centrosymmetric (with an odd number of unit cells in each dimension)\cite{Zhao_Hu2025,Zhao_Hu2026,He_Hu2026}. 

For each parallelogram ${\mathbf S}_j$ spanned by vectors ${\mathbf a}_j$ and ${\mathbf b}_j$, the outward-pointing normal vector is proportional to ${\mathbf a}_j\times {\mathbf b}_j$.
The differential surface vector element is given by
\begin{equation} d{\mathbf S}_j = \left( {\mathbf a}_j\times {\mathbf b}_j \right) dt d\tau , \end{equation}
and any point ${\mathbf x}$ on this facet is parameterized as
\begin{equation} {\mathbf x} = t{\mathbf a}_j + \tau {\mathbf b}_j + {\mathbf c}_j, \quad \text{for }\, t,\tau \in [-1,1], \end{equation}
where ${\mathbf c}_j$ is the position vector from the center of the crystal to the center of the $j$-th parallelogram.

The second step involves evaluating this parameterized integral by recognizing that the gradient with respect to ${\mathbf x}$ equals the gradient with respect to ${\mathbf c}_j$
\begin{equation} \nabla_{\mathbf x}\frac{1}{\left| {\mathbf x} \right|} = -\frac{\mathbf x}{ \left| {\mathbf x} \right|^3} = \nabla_{{\mathbf c}_j} \frac{1}{\left| t{\mathbf a}_j + \tau {\mathbf b}_j + {\mathbf c}_j \right|}.  \end{equation}
This identity allows us to perform the integration over the inverse distance first for each facet, followed by the gradient operation with respect to ${\mathbf c}_j$.
Consequently, the boundary term can be expressed as a sum over the crystal facets:
\begin{equation} \nu_b({\mathbf r}|{\mathbf s}) = \frac{1}{2V} \sum_{j=1}^{N_{\mathbf s}}\frac{{\mathbf r}\cdot\left({\mathbf a}_j\times {\mathbf b}_j\right)}{\left| {\mathbf a}_j\times {\mathbf b}_j\right|} {\mathbf r}\cdot {\mathbf g}({\mathbf a}_j,{\mathbf b}_j, {\mathbf c}_j) ,\label{eq:nub} \end{equation}
where the vector field ${\mathbf g}$ is defined as the gradient of the electrostatic potential,
\begin{equation} {\mathbf g}({\mathbf a},{\mathbf b},{\mathbf r}) = \nabla \phi({\mathbf a},{\mathbf b},{\mathbf r}) ,\end{equation}
and $\phi$ represents the electrostatic potential of a uniformly charged parallelogram, given by the surface integral of the inverse distance:
\begin{equation} \phi({\mathbf a},{\mathbf b},{\mathbf r}) = \iint_{-1}^1 \frac{dt\,d\tau \left|{\mathbf a}\times {\mathbf b}\right| }{\left| t {\mathbf a} + \tau {\mathbf b} + {\mathbf r} \right|} =
 \iint_{-1}^1 \frac{dt\, d\tau \left|{\mathbf a}\times {\mathbf b}\right| }{\left| t {\mathbf a} + \tau {\mathbf b} - {\mathbf r} \right|} .\end{equation}
This integral corresponds to the electrostatic potential at a field point ${\mathbf r}$ arising from a uniformly charged parallelogram centered at the origin. 
The explicit evaluation of this integral via Euler substitution is provided in the Appendices A and B.

%%%%%%%%%%%%%%%%%%%%%%%%%%%%%%%%%%%%%%%%%%%%%%%%%%%%%%%%%%%%%%%%%%%%%%%%%%%%%%%%%%%%%%%%%%%%%%%%%%%%%%%%%%%%%%%%%%%%%%%%%%%%%%%%%%%%%%%%%%%%%%%%%%%%%%%%%%%%%%%%
\section{Analytical Expression of the Boundary Term for Arbitrary Crystal Geometries}\label{sec:result}
We now present the explicit expressions for both $\phi$ and ${\mathbf g}$, followed by applications of Eq.~\eqref{eq:nub} to representative crystals.
The electrostatic potential $\phi({\mathbf a},{\mathbf b},{\mathbf r})$ depends on three fundamental vectors: the spanning vectors ${\mathbf a}$ and ${\mathbf b}$ of the parallelogram, and the field point ${\mathbf r}$.
To evaluate $\phi$ analytically, we introduce a set of auxiliary geometric variables and scalar functions. 
First, the magnitudes and unit vectors of ${\mathbf a}$ and ${\mathbf b}$ are given by
\begin{equation} 
a = \left| {\mathbf a} \right|; \quad b = \left| {\mathbf b} \right|; \quad {\mathbf e}_a = {\mathbf a}/a; \quad {\mathbf e}_b = {\mathbf b}/b.   \label{eq:ab}
\end{equation}
Next, we define a set of basis vectors and their common scalar magnitude $v$
\begin{align} 
{\mathbf v}_1 &= {\mathbf e}_a - {\mathbf e}_b \left( {\mathbf e}_a\cdot {\mathbf e}_b \right) \equiv {\mathbf e}_b\times \left( {\mathbf e}_a \times {\mathbf e}_b \right),  \\ 
{\mathbf v}_2 &= {\mathbf e}_b - {\mathbf e}_a \left( {\mathbf e}_a\cdot {\mathbf e}_b \right) \equiv {\mathbf e}_a\times \left( {\mathbf e}_b \times {\mathbf e}_a \right), \\ 
{\mathbf v}_0 &= {\mathbf e}_a \times {\mathbf e}_b; \quad v = \left| {\mathbf v}_0 \right| = \left| {\mathbf v}_1 \right| = \left| {\mathbf v}_2 \right|.  
\end{align}
The scaled projections of the field point ${\mathbf r}$ onto these basis vectors are defined as
\begin{equation} 
r_0 = {\mathbf r}\cdot \frac{{\mathbf v}_0}{v};\quad r_1 = {\mathbf r}\cdot \frac{{\mathbf v}_1}{v}; \quad r_2 = {\mathbf r}\cdot \frac{{\mathbf v}_2}{v}.  
\end{equation}
The relative position vectors from the parallelogram vertices ($t,\tau=\pm 1$) to the field point, along with their magnitudes, are given by
\begin{equation} 
{\mathbf r}_s(t,\tau) = {\mathbf r} + t {\mathbf a} - \tau {\mathbf b};\quad r_s(t,\tau) = \left| {\mathbf r}_s(t,\tau) \right|,  
\end{equation}
where the indices $t,\tau \in \{-1, +1\}$. Finally, the auxiliary scalar functions are defined as
\begin{align}
& s_1(t,\tau) = r_s(t,\tau) + {\mathbf e}_a \cdot {\mathbf r}_s(t,\tau), \\
& s_2(t,\tau) = r_s(t,\tau) - {\mathbf e}_b \cdot {\mathbf r}_s(t,\tau), \\
& s_3(t,\tau) = r_s(t,\tau)\left(1-{\mathbf e}_a\cdot{\mathbf e}_b\right) + {\mathbf r}_s(t,\tau)\cdot\left( {\mathbf e}_a - {\mathbf e}_b \right). \label{eq:s3}
\end{align}

With the above parameters defined, the potential $\phi({\mathbf a},{\mathbf b},{\mathbf r})$ decomposes into logarithmic (log) and arctangent (atan) contributions
\begin{equation} 
\phi({\mathbf a},{\mathbf b},{\mathbf r}) = \phi_{\rm log}({\mathbf a},{\mathbf b},{\mathbf r}) + \phi_{\rm atan}({\mathbf a},{\mathbf b},{\mathbf r}),  \label{eq:phi}
\end{equation}
where the logarithmic and arctangent contributions are given, respectively, by
\begin{multline} 
\phi_{\rm log}({\mathbf a},{\mathbf b},{\mathbf r}) = \sum_{t,\tau=\pm 1} \tau\left(av + t r_1\right) \ln s_2(t,\tau) \\ 
+ \sum_{t,\tau=\pm 1} t\left(bv - \tau r_2\right) \ln s_1(t,\tau), 
\end{multline}
and
\begin{equation} 
\phi_{\rm atan}({\mathbf a},{\mathbf b},{\mathbf r}) = 2r_0 \sum_{t,\tau=\pm 1} t \tau\, {\rm atan}\left(\frac{s_3(t,\tau)}{v r_0}\right).  \label{eq:phia}
\end{equation}

To compute the gradient of $\phi$ with respect to the field point ${\mathbf r}$, which is required to evaluate the boundary term, we introduce the following auxiliary vector functions:
\begin{align}
 {\mathbf d}(t,\tau) & =\nabla r_s(t,\tau)  = {\mathbf r}_s(t,\tau)/r_s(t,\tau), \\ % \frac{{\mathbf r} + t {\mathbf a} - \tau {\mathbf b}}{\left| {\mathbf r} + t {\mathbf a} - \tau {\mathbf b} \right|}  \end{equation}
{\mathbf d}_1(t,\tau)& = \nabla s_1(t,\tau) = {\mathbf d}(t,\tau) + {\mathbf e}_a, \\
{\mathbf d}_2(t,\tau)& = \nabla s_2(t,\tau) = {\mathbf d}(t,\tau) - {\mathbf e}_b,
\end{align}
and
\begin{equation} {\mathbf d}_3(t,\tau) = \nabla s_3(t,\tau) = \left( 1- {\mathbf e}_a \cdot {\mathbf e}_b \right) {\mathbf d}(t,\tau) + {\mathbf e}_a - {\mathbf e}_b . \end{equation}

The total gradient ${\mathbf g}({\mathbf a},{\mathbf b},{\mathbf r}) = \nabla_{\mathbf r}\phi({\mathbf a},{\mathbf b},{\mathbf r})$ decomposes into five distinct vector components: three algebraic terms (${\mathbf G}_1$, ${\mathbf G}_2$, ${\mathbf G}_3$), a logarithmic term (${\mathbf L}$), and an arctangent term (${\mathbf A}$):
\begin{equation} 
{\mathbf g}({\mathbf a},{\mathbf b},{\mathbf r}) = {\mathbf G}_1 + {\mathbf G}_2 + {\mathbf G}_3 + {\mathbf L} + {\mathbf A}, 
\end{equation}
where
\begin{equation} {\mathbf G}_1({\mathbf a},{\mathbf b},{\mathbf r}) = \sum_{t,\tau=\pm 1} t\left(bv - \tau r_2\right) \frac{{\mathbf d}_1(t,\tau)}{s_1(t,\tau)}, \end{equation}
\begin{equation} {\mathbf G}_2({\mathbf a},{\mathbf b},{\mathbf r}) = \sum_{t,\tau=\pm 1} \tau\left(av + t r_1\right) \frac{{\mathbf d}_2(t,\tau)}{s_2(t,\tau)}, \end{equation}
\begin{equation} {\mathbf G}_3({\mathbf a},{\mathbf b},{\mathbf r}) = 2r_0 \sum_{t,\tau=\pm 1} t\tau \frac{v r_0 {\mathbf d}_3(t,\tau) - s_3(t,\tau) {\mathbf v}_0 }{v^2 r_0^2+s_3^2(t,\tau)}, \end{equation}
\begin{equation} {\mathbf L}({\mathbf a},{\mathbf b},{\mathbf r}) = \sum_{t,\tau=\pm 1} t \tau \frac{{\mathbf v}_1 \ln s_2(t,\tau) - {\mathbf v}_2 \ln s_1(t,\tau)}{v}, \end{equation}
and 
\begin{equation} {\mathbf A}({\mathbf a},{\mathbf b},{\mathbf r}) = \frac{2 {\mathbf v}_0}{v} \sum_{t,\tau=\pm 1} t \tau\, \arctan\left(\frac{s_3(t,\tau)}{v r_0}\right). \end{equation}

The vector field ${\mathbf g}({\mathbf a}, {\mathbf b}, {\mathbf r})$ exhibits specific symmetry properties: it is invariant under the exchange of the spanning vectors ${\mathbf a}$ and ${\mathbf b}$, 
even with respect to the sign reversal of either ${\mathbf a}$ or ${\mathbf b}$, and odd with respect to the field vector ${\mathbf r}$; i.e.,
\begin{align} 
{\mathbf g}({\mathbf a}, {\mathbf b}, {\mathbf r}) &= {\mathbf g}({\mathbf b}, {\mathbf a}, {\mathbf r}), \\
{\mathbf g}({\mathbf a}, {\mathbf b}, {\mathbf r}) &= {\mathbf g}(\pm{\mathbf a}, \pm{\mathbf b}, {\mathbf r}), \\
{\mathbf g}({\mathbf a}, {\mathbf b}, {\mathbf r}) &= -{\mathbf g}({\mathbf a}, {\mathbf b}, -{\mathbf r}). \label{eq:gr}
\end{align}
These symmetries provide useful geometric flexibility when evaluating Eq.~\eqref{eq:nub}. 
Specifically, if the spanning vectors ${\mathbf a}_j$ and ${\mathbf b}_j$ are chosen such that their cross product ${\mathbf a}_j \times {\mathbf b}_j$ points inward (toward the crystal interior) rather than outward, the boundary term remains strictly invariant, 
provided ${\mathbf c}_j$ is simultaneously redefined as the vector pointing from the center of the parallelogram to the center of the crystal (i.e., ${\mathbf c}_j \to -{\mathbf c}_j$). 
The resulting sign flips in both the surface normal (the prefactor in Eq.~\eqref{eq:nub}) and the third argument of ${\mathbf g}$ [via Eq.~\eqref{eq:gr}] exactly cancel one another.
Owing to these symmetries, it suffices to specify only $N_{\mathbf s}/2$ parallelograms.
Furthermore, these facets are generated by permuting just $N_{\mathbf s}/6$ fundamental sets of vectors.
Representative fundamental sets of vectors ${\mathbf a}_j$, ${\mathbf b}_j$, and ${\mathbf c}_j$ corresponding to the shapes in Fig.~\ref{fig:tshape} are listed in Table~\ref{tab:abc}.
Because the boundary term is invariant under uniform scaling, the fundamental vectors in Table~\ref{tab:abc} are expressed in terms of the primitive lattice vectors with coefficients proportional to the crystal size.
\begin{table}[!htb]   %%%%%%%%%%%%%%%%%%%%%%%%%%%%%%%%%%%%%%%%  long table ---------------------------------------------------unit cell coordinates---------------
\caption{Representative sets of vectors defining the boundary term for the regular (first row) and irregular (remaining $5$ rows) crystal shapes shown in Fig.~\ref{fig:tshape}.}
\label{tab:abc}
\begin{tabular}{ccc}\hline    $\quad{\mathbf a}_j\quad$     &   $\quad{\mathbf b}_j\quad$    &      ${\mathbf c}_j$     \\[0.5ex] \hline
                                  ${\mathbf u}_1$     &      ${\mathbf u}_2$     &      ${\mathbf u}_3$       \\[0.5ex] \hline
                                  ${\mathbf u}_1$     &      ${\mathbf u}_2$     &      $3{\mathbf u}_3$       \\[0.5ex]
                                  ${\mathbf u}_2$     &      ${\mathbf u}_3$     &      $2{\mathbf u}_3+{\mathbf u}_1$       \\[0.5ex]
                                  ${\mathbf u}_3$     &      ${\mathbf u}_2$     &      $2{\mathbf u}_3-{\mathbf u}_1$       \\[0.5ex]
                                  ${\mathbf u}_3$     &      ${\mathbf u}_1$     &      $2{\mathbf u}_3+{\mathbf u}_2$       \\[0.5ex]
                                  ${\mathbf u}_1$     &      ${\mathbf u}_3$     &      $2{\mathbf u}_3-{\mathbf u}_2$       \\ \hline
\end{tabular} \end{table}

\begin{table*}[!htb]   %%%%%%%%%%%%%%%%%%%%%%%%%%%%%%%%%%%%%%%%
\caption{Madelung constants for NaCl (${\mathbf r}_1 = (1,1,1)$) and ZnS (${\mathbf r}_2 = (0.5, 0.5, 0.5)$), along with their absolute errors, as a function of the crystal size $p$ ($K=2p+1$). 
The finite crystals adopt the regular shape shown in the left panel of Fig.~\ref{fig:tshape}. Exact values are $1.74756459463318$ for NaCl\cite{naclseq} and $1.6380550533887894$ for ZnS\cite{znsseq}.}\label{tab:naclzns}
\begin{tabular}{cccccccc}\hline p& $\nu-\nu_{\rm b}$ & $\nu - \nu_{\rm b}-\nu_{\rm corr}$ & $\epsilon$            &$\left(\nu-\nu_{\rm b}\right)\sqrt{3}/2$ &  $\left(\nu - \nu_{\rm b}-\nu_{\rm corr}\right)\sqrt{3}/2$ &   $\epsilon$         \\[0.4ex] \hline
                            1   &   1.771344         &  1.73686204609                     &  -1.1$\times 10^{-2}$ &               1.632246                  &  1.65933993370     &\,2.1$\times 10^{-2}$  \\[0.5ex]
                            2   &   1.756833         &  1.74867033258                     & \,1.1$\times 10^{-3}$ &               1.635908                  &  1.63796699572     & -8.8$\times 10^{-5}$ \\[0.5ex]
                            3   &   1.752284         &  1.74754626064                     &  -1.8$\times 10^{-5}$ &               1.636954                  &  1.63804491006     & -1.0$\times 10^{-5}$ \\[0.5ex]
                           10   &   1.748088         &  1.74756445162                     &  -1.4$\times 10^{-7}$ &               1.637932                  &  1.63805499708     & -5.6$\times 10^{-8}$ \\[0.5ex]
                           20   &   1.747702         &  1.74756458566                     &  -9.0$\times 10^{-9}$ &               1.638023                  &  1.63805504995     & -3.4$\times 10^{-9}$ \\[0.5ex]
                           60   &   1.747580         &  1.74756459452                     &  -1.1$\times 10^{-10}$&               1.638051                  &  1.63805505335     & -4.0$\times 10^{-11}$\\ \hline        
\end{tabular} \end{table*}
\begin{table*}[!htb]   %%%%%%%%%%%%%%%%%%%%%%%%%%%%%%%%%%%%%%%%
\caption{Same as Table~\ref{tab:naclzns}, but for finite crystals with the irregular shape shown in the right panel of Fig.~\ref{fig:tshape}.}\label{tab:ir}
\begin{tabular}{cccccccc}\hline p& $\nu-\nu_{\rm b}$ & $\nu - \nu_{\rm b}-\nu_{\rm corr}$ & $\epsilon$            &$\left(\nu-\nu_{\rm b}\right)\sqrt{3}/2$ &  $\left(\nu - \nu_{\rm b}-\nu_{\rm corr}\right)\sqrt{3}/2$ &   $\epsilon$         \\[0.4ex] \hline
                            1   &   1.736838         &  1.72223272215                    &  -2.5$\times 10^{-2}$ &               1.644976                  &  1.63719597450     & -8.6$\times 10^{-4}$  \\[0.5ex]
                            2   &   1.743602         &  1.74740610587                    &  -1.6$\times 10^{-4}$ &               1.640573                  &  1.63809597799     & \,4.1$\times 10^{-5}$ \\[0.5ex]
                            3   &   1.745540         &  1.74755980831                    &  -4.8$\times 10^{-6}$ &               1.639343                  &  1.63806280573     & \,7.8$\times 10^{-6}$ \\[0.5ex]
                           10   &   1.747340         &  1.74756457145                    &  -2.3$\times 10^{-8}$ &               1.638199                  &  1.63805509614     & \,4.3$\times 10^{-8}$ \\[0.5ex]
                           20   &   1.747506         &  1.74756459321                    &  -1.4$\times 10^{-9}$ &               1.638093                  &  1.63805505598     & \,2.6$\times 10^{-9}$ \\[0.5ex]
                           60   &   1.747558         &  1.74756459462                    &  -1.6$\times 10^{-11}$&               1.638059                  &  1.63805505344     & \,4.7$\times 10^{-11}$\\ \hline
\end{tabular} \end{table*}

In the special case of a rectangular prism, the input vectors ${\mathbf a}$, ${\mathbf b}$, and ${\mathbf r}$ are mutually orthogonal.
Consequently, the distance $r_s(t,\tau)$ becomes independent of the indices $t$ and $\tau$. 
By symmetry, the logarithmic term ${\mathbf L}$ vanishes, as do the components of ${\mathbf G}_1$, ${\mathbf G}_2$, and ${\mathbf G}_3$ perpendicular to ${\mathbf r}$.
Furthermore, the sum of the algebraic terms ${\mathbf G}_1+{\mathbf G}_2 + {\mathbf G}_3$ exactly cancels in the direction parallel to ${\mathbf r}$, leaving only the arctangent term ${\mathbf A}$.
By aligning the vectors along the Cartesian axes, e.g., ${\mathbf a}=(\xi_1,0,0)$, ${\mathbf b}=(0,\xi_2,0)$, and ${\mathbf r}=(0,0,\xi_3)$, the arctangent term simplifies via the identity in Eq.~\eqref{eq:iarc} of the Appendix:
\begin{equation} {\mathbf g}({\mathbf a}, {\mathbf b}, {\mathbf r}) = -4\, {\rm atan}\frac{\xi_2\xi_3}{\xi_1\sqrt{\xi_1^2+\xi_2^2+\xi_3^2}}.  \end{equation}
Consequently, the boundary term for a rectangular prism with side lengths $\xi_1$, $\xi_2$, and $\xi_3$ is given by
\begin{equation} \nu_{\rm b}({\mathbf r}|{\mathbf s}) = -\frac{4}{V} \sum_{j=1}^3 \left({\mathbf r}\cdot{\mathbf e}_j\right)^2 \,{\rm atan}\frac{\xi_1\xi_2\xi_3}{\xi_j\sqrt{\xi_1^2+\xi_2^2+\xi_3^2}} , \label{eq:br} \end{equation}
where ${\mathbf e}_j$ is the Cartesian unit vectors.
Evidently, the boundary term depends exclusively on the dimensionless ratios $\xi_1:\xi_2:\xi_3$, rather than on the overall crystal volume $\xi_1\xi_2\xi_3$.
The required ratios are subsequently obtained from the shape parameters ${\mathbf s}$ and the lattice parameters of the unit cell.
Nevertheless, Eq.~\eqref{eq:br} is in agreement with earlier derivations\cite{Pan2017,Zhao_Hu2026}.
%%%%%%%%%%%%%%%%%%%%%%%%%%%%%%%%%%%%%%%%%%%%%%%%%%%%%%%%%%%%%%%%%%%%%%%%%%%%%%%%%%%%%%%%%%%%%%%%%%%%%%%%%%%%%%%%%%%%%%%
\section{Properties of a composite medium with obstacles arranged in square order}\label{sec:ray}

\begin{figure}[!htb] \centerline{\includegraphics[width=7cm]{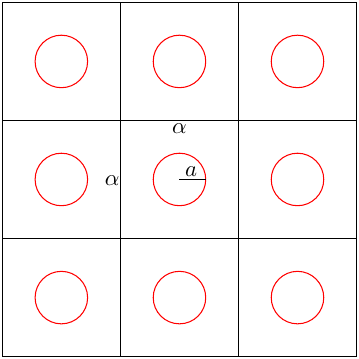}}           %---------------------  Figure -------------------------------------
\caption{
Cylindrical or spherical obstacles of radius $a$ arranged in a square or cubic lattice, with each obstacle occupying a unit cell of side $\alpha$. 
The macroscopic current (or applied field) is directed along the $x$-axis, i.e., in the $(1,0,0)$ direction.
The cylinders are assumed to extend infinitely in the perpendicular direction.
} \label{fig:obs} \end{figure}

So far, we have focused on the Madelung series and obtained a physical explanation for the boundary term. 
This mathematical framework is of broad interest and finds a classic application in the foundational work of Lord Rayleigh\cite{Rayleigh1892}, who investigated the influence of periodic obstacles on the transport properties of a medium.
Rayleigh derived analytical results for both two-dimensional cylindrical and three-dimensional spherical obstacles arranged in regular lattices, as shown in fig.~\ref{fig:obs}.

For the case of electric conductivity with cylindrical obstacles and current flowing in the $x$-direction, the relative effective conductivity $\sigma$ of the composite medium is expressed as 
\begin{equation} \sigma = 1 - \frac{2\pi B_1}{\alpha^2 H},  \label{eq:sigma} \end{equation}
where $H$ is the applied macroscopic field gradient, and the ratio $H/B_1$ is expressed as a series expansion involving lattice sums:
\begin{equation} 
    \frac{H a^2}{B_1} = \frac{1+v}{1-v} + \frac{a^2}{\alpha^2} S_2 - \frac{3(1-v)}{1+v} \frac{a^8}{\alpha^8} S_4^2 + \cdots. 
\end{equation}
Here, $v$ denotes the ratio of the conductivity of the material composing the cylinders to that of the background medium. 
The lattice sum $S_2$ is conditionally convergent and depends on the macroscopic boundary shape of the sample, whereas the higher-order series possess no such conditional convergence and can be evaluated straightforwardly.

Similarly, for three-dimensional spherical obstacles, the relative conductivity is given by
\begin{equation}  \sigma = 1 - \frac{4\pi B_1}{\alpha^3 H},  \label{eq:sigma2} \end{equation}
where the corresponding expansion is
\begin{equation} \frac{H a^3}{B_1} = \frac{2+v}{1-v} + \frac{2a^3}{\alpha^3} S_2 - \frac{96}{5}\frac{1-v}{4+3v} \frac{a^{10}}{\alpha^{10}} S_4^2 + \cdots.  \end{equation}

The conditional convergence of the $S_2$ lattice sum is intimately tied to the macroscopic shape of the sample, which manifests mathematically as a boundary term. 
By applying our generalized framework to these specific geometries, we can exactly recover Rayleigh's geometric factors. 
Specifically, for the 2D cylindrical case, taking the limit of an infinite prism by choosing $\xi_1=\xi_2$ and $\xi_3=\infty$, with the field direction ${\mathbf r}=(1,0,0)$ and a unit cell volume of unity in Eq.~\eqref{eq:br}, yields a boundary term of exactly $-2\pi$. Conversely, for the 3D spherical case, choosing an isotropic limit $\xi_1 = \xi_2 = \xi_3$ yields a boundary term of exactly $-2\pi/3$. 
These factors of $2\pi$ and $2\pi/3$ emerge naturally in our formalism, perfectly mirroring the geometric coefficients derived by Rayleigh through his approach using direct integration.

Furthermore, as Rayleigh explicitly emphasized in his original work, the refractive index and other transport properties, such as heat conduction, can be treated using a mathematical framework identical to that of electrical conductivity in Eqs.~\eqref{eq:sigma} and~\eqref{eq:sigma2}. 
The resulting expression for the effective refractive index naturally generalizes the classical Lorenz-Lorentz relation. 
These diverse physical contexts highlight the broad applicability of the conditional boundary term. 
Consequently, our generalized formulation for arbitrary macroscopic sample shapes provides a robust framework to extend and clarify Rayleigh's earlier results across these various domains.

\section{Applications to various ionic crystals}\label{sec:app}

By isolating the bulk contribution from the finite-lattice sum, we obtain
\begin{equation} \nu_{\rm pbc}({\mathbf r}) = \nu({\mathbf r},p|{\mathbf s}) - \nu_{\rm b}({\mathbf r}|{\mathbf s}) - \nu_{\rm corr}({\mathbf r},p|{\mathbf s}), \label{eq:pbc} \end{equation}
which establishes an explicitly corrected direct-summation scheme for evaluating the bulk potential.
While integral-transform techniques --- most notably Ewald summation\cite{Ewald1921,DeLeeuw_Smith1980}, its variants\cite{Nijboer1957,Borwein1985,Darden1993,Essmann1995,Petersen1995,Gao_Greengard2026} and fast multipole methods\cite{Greengard1987,Greengard1988,Kudin_Scuseria1998} --- are highly accurate and widely used, direct-summation approaches\cite{Evjen1932,Harrison2006,Marathe1983,Sousa1993,Derenzo2000,Gelle2008,Tavernier2020,Tavernier2021} remain conceptually simpler and physically more transparent.
Previously, highly accurate results were obtained for cubic crystal systems using larger unit cells containing multiple formula units.
Here, we demonstrate the correctness of our formulation by computing Madelung constants for several representative structures using the smallest possible unit cell (single formula unit) of an FCC Bravais lattice.

Specifically, we consider the rocksalt (NaCl), zincblende (ZnS), and fluorite (CaF$_2$) structures.
The primitive vectors of these structures in the FCC Bravais lattice are the same: ${\mathbf u}_1 = (0,1,1)$, ${\mathbf u}_2 = (1,0,1)$, and ${\mathbf u}_3 = (1,1,0)$. 
With the cations (Na$^+$, Zn$^{2+}$, and Ca$^{2+}$) placed at the origin, the reduced coordinates for the anions are: $(1,0,0)$ for Cl$^-$, $(0.5, 0.5, 0.5)$ for S$^{2-}$/F$^-$, and $(-0.5, -0.5, -0.5)$ for the other F$^-$ in CaF$_2$.
Using the boundary-term formulation developed above, we compute $\nu_{\rm pbc}$ with an accuracy of $\mathcal{O}(p^{-2})$.
The leading-order finite-size correction can be estimated from the difference $\nu({\mathbf r},p|{\mathbf s}) - \nu({\mathbf r},p-1|{\mathbf s})$, which scales as $C\left[ (2p+1)^{-2} - (2p-1)^{-2} \right]$.
From the coefficient $C$, we obtain the finite-size correction term $C/(2p+1)^2$.
Incorporating this correction significantly improves the numerical accuracy, as demonstrated in Table~\ref{tab:naclzns}.
For CaF$_2$, the cation occupies the same local environment as in ZnS, with two distinct displacement vectors ${\mathbf r}_1$ and ${\mathbf r}_2$ for the two fluorine ions.
The electrostatic potential at each F$^-$ site is a linear combination of the contributions from these two vectors, and comparable accuracy is therefore expected.

The same analysis can be applied to irregular crystal shapes, such as the one shown in the right panel of Fig.~\ref{fig:tshape}, whose fundamental sets of vectors are listed in Table~\ref{tab:abc}.
Although the boundary contributions $\nu_{\rm b}$ differ for irregular shapes, the final corrected results converge to comparable accuracy, as demonstrated in Table~\ref{tab:ir}.
This consistency across different crystal shapes confirms the correctness and robustness of our analytical formulation.
The numerical results presented in Tables~\ref{tab:naclzns} and \ref{tab:ir} confirm that our formulation correctly isolates the bulk potential $\nu_{\rm pbc}$ from the finite-lattice sum, achieving convergence to the known exact Madelung constants with high precision.
The systematic improvement with increasing $p$, together with the agreement between regular and irregular shapes, collectively validates the analytical boundary-term expressions derived in this work.

\begin{table}[!htb]   %%%%%%%%%%%%%%%%%%%%%%%%%%%%%%%%%%%%%%%%
\caption{Electrostatic energies (in e$^2$/\AA) of ions in the triclinic wollastonite lattice computed at $p=20$, accurate to six digits, compared with those of the cubic perovskite structure (last row, unit cell length $l=4.04740219773$\AA).}\label{tab:casio3}
\begin{tabular}{c|c|ccc}\hline  Ca$^{2+}$ & Si$^{4+}$ &          & O$^{2-}$ &           \\[0.4ex] \hline
                               -1.54032   & -6.65834  & -1.74925 & -1.80131 & -2.12904  \\[0.5ex]
                               -1.48905   & -6.63164  & -1.73140 & -1.79153 & -2.14650  \\[0.5ex]
                               -1.51901   & -6.62814  & -1.78680 & -1.83081 & -2.12679  \\[0.5ex] \hline
                               -1.33103   & -6.11625  &          & -1.59507 &           \\ \hline
\end{tabular} \end{table}
We now proceed to compute the bulk potential of a triclinic lattice, specifically wollastonite (CaSiO$_3$)\cite{Buerger_Prewitt1961,wollastonite}, and compare it with that of the usual cubic perovskite-type structure of the same density.
The unit cell of this triclinic lattice is characterized by lattice parameters $a = 7.94$~\AA, $b = 7.32$~\AA, $c = 7.07$~\AA, and angles $\alpha = 90.033^\circ$, $\beta = 95.367^\circ$, and $\gamma = 103.433^\circ$\cite{Buerger_Prewitt1961}.
The conversion of these parameters to primitive vectors follows the standard procedure\cite{wollastonite}.
Each unit cell contains six formula units of CaSiO$_3$, possesses inversion symmetry, and consequently has a vanishing total dipole moment: $\sum_{j=1}^M q_j = 0$.

Under the conditions that both the total charge and the dipole moment vanish, the contribution of the boundary term to the electrostatic potential becomes a constant for all charges:
\begin{equation} \sum_{j=1}^N q_j \nu_{\rm b}({\mathbf r}_j - {\mathbf r}_i | {\mathbf s}) = \sum_{j=1}^N q_j \nu_{\rm b}({\mathbf r}_j | {\mathbf s}), \label{eq:nuphi} \end{equation}
where $\nu_{\rm b}({\mathbf r}|{\mathbf s})$ is explicitly given by Eq.~\eqref{eq:nub}.
Using the atomic coordinates of wollastonite\cite{Buerger_Prewitt1961,wollastonite}, Eq.~\eqref{eq:nuphi} yields a constant value of $0.04736$ e/\AA.
This constant contribution to the potential, however, does not affect the total electrostatic energy of the unit cell, as its net contribution vanishes.
Using a relatively small crystal of regular shape at $p=20$ (see Fig.~\ref{fig:tshape}), highly accurate electrostatic energies are obtained, as shown in Table~\ref{tab:casio3}.
For the cubic perovskite-type structure at the same density, the cubic unit cell has a lattice parameter $l = 4.04740219773$~\AA, and the corresponding electrostatic energies for each ion are also listed in Table~\ref{tab:casio3}.
The electrostatic energies of all ions are clearly lowered relative to those of the cubic perovskite structure, thereby favoring the cubic-to-triclinic phase transition.

%%%%%%%%%%%%%%%%%%%%%%%%%%%%%%%%%%%%%%%%%%%%%%%%%%%%%%%%%%%%%%%%%%%%%%%%%%%%%%%%%%%%%%%%%%%%%%%%%%%%%%%%%%%%%%%%%%%%%%%%%%%%%%%%%%%%%%%%%%%%%%%%%%%%%%%%%%%%%%%%
\section{Conclusion}\label{sec:con}
To summarize, we have derived a closed-form analytical expression for the shape-dependent boundary term in lattice sums that is valid for arbitrary crystal geometries.
The derivation resolves the challenging problem left from the seminal work of De Leeuw, Perram, and Smith\cite{DeLeeuw_Smith1980,Smith1981}.
The explicit evaluation of this boundary term, together with the systematic finite-size correction that decays as $(2p+1)^{-2}$, provides a robust and physically transparent direct-summation framework for computing Madelung constants. 
Numerical validation on NaCl, ZnS, and CaF$_2$ confirms convergence to known exact values with high precision, while the application to triclinic wollastonite reveals a quantitative energetic driving force for the cubic-to-triclinic phase transition. 
By resolving a long-standing analytical challenge, this work offers a universally applicable tool for accurate electrostatic calculations in complex low-symmetry condensed matter systems.

\section*{Acknowledgement}
This work was supported by NSFC (Grant No. 22273047) and Shandong Provincial Special Zone for Fundamental Research (Chemistry) (Grant No. TQ022025003).
%%%%%%%%%%%%%%%%%%%%%%%%%%%%%%%%%%%%%%%%%%%%%%%%%%%%%%%%%%%%%%%%%%%%%%%%%%%%%%%%%%%%%%%%%%%%%%%%%%%%%%%%%%%%%%%%%%%%%%%%%%%%%%%%%%%%%%%%%%%%%%%%%%%%%%%%%%%%%%%%
%\newpage \,\quad\, \newpage
\section{Appendix A:  Evaluation the integration of a logarithmic-radical expression via Euler substitution}
\renewcommand{\theequation}{A\arabic{equation}}  % ← key line
\setcounter{equation}{0}                          % restart numbering
To evaluate the indefinite integral
\begin{equation}  I = \int du\, \ln T(u) ; \quad T(u) = \sqrt{u^2 + \alpha^2} + \beta u + \gamma, \label{eq:alog} \end{equation}
we begin with integration by parts:
\begin{equation}     \int du\, \ln T(u) = u \ln T(u) - \int du\, u T'(u)/T(u) . \label{eq:ipart} \end{equation}
Since $T'(u) = \beta + u/\sqrt{u^2+\alpha^2}$, we compute
\begin{equation} uT'(u) = \frac{u^2}{\sqrt{u^2+\alpha^2}} + \beta u = T(u) - \gamma - \frac{\alpha^2}{\sqrt{u^2+\alpha^2}}. \end{equation}
Dividing by $T(u)$ and integrating yields
\begin{equation} \int u \frac{T'(u)}{T(u)} \, du = u - \int \frac{\gamma \,du}{T(u)} - \int \frac{\alpha^2 \,du}{\sqrt{u^2+\alpha^2}\, T(u)}. \end{equation}
Substituting back into Eq.~\eqref{eq:ipart}, we obtain
\begin{equation}     I = u \ln T(u) - u + \gamma I_1 + \alpha^2 I_2 + C, \label{eq:ibp_result} \end{equation}
where $C$ is an integration constant, and we define
\begin{equation}    I_1 = \int \frac{du}{T(u)}; \qquad I_2 = \int \frac{du}{\sqrt{u^2+\alpha^2}\, T(u)}. \label{eq:i1i2} \end{equation}

To evaluate $I_1$ and $I_2$, we apply the Euler substitution
\begin{equation}  t = \sqrt{u^2 + \alpha^2} + u, \label{eq:euler_sub}\end{equation}
which is strictly increasing for $\alpha \neq 0$.
Solving for $u$ and its differential gives
\begin{equation}  u = \frac{t^2 - \alpha^2}{2t}; \,  \sqrt{u^2 + \alpha^2} = \frac{t^2 + \alpha^2}{2t}; \, \frac{du}{dt} = \frac{t^2 + \alpha^2}{2t^2}. \end{equation}
Under this substitution, $T(u)$ simplifies to
\begin{equation}   T(u) = \frac{(1+\beta)t^2 + 2\gamma t + (1-\beta)\alpha^2}{2t} \equiv \frac{R(t)}{2t}, \end{equation}
where $R(t) = (1+\beta)t^2 + 2\gamma t + (1-\beta)\alpha^2$ is a quadratic polynomial. 
The integrals $I_1$ and $I_2$ transform as
\begin{align} \frac{du}{T(u)} &= \frac{t^2 + \alpha^2}{t R(t)} \, dt   \nonumber \\
                              &= \frac{1}{1-\beta}\frac{dt}{t} - \frac{\beta\, dR(t)}{\left(1-\beta^2\right) R(t)} - \frac{2\gamma}{1-\beta^2}\frac{dt}{R(t)}, \label{eq:duT} \end{align}
and
\begin{equation} \frac{du}{\sqrt{u^2+\alpha^2} T(u)} = 2 \frac{dt}{R(t)}.   \end{equation}
Here, both expressions have been further decomposed into partial fractions, converting $I_1$ and $I_2$ into integrals of rational functions. 
Completing the square in $R(t)$ yields
\begin{equation}  R(t) = (1+\beta)\left[ \left(t + \frac{\gamma}{1+\beta}\right)^2 + \frac{1-\beta}{1+\beta} K \right], \end{equation}
where the discriminant-like parameter $K$ is given by
\begin{equation} K = \alpha^2 - \gamma^2/\left(1-\beta^2\right). \label{eq:k} \end{equation}
When $K > 0$ and $ 1 - \beta^2 > 0 $, the polynomial $R(t)$ is positive definite, and the remaining integrals in Eq.~\eqref{eq:i1i2} evaluate to elementary logarithmic and arctangent functions.
The final result for the indefinite integral reads
\begin{multline} I = u\ln T(u) -u + \frac{\gamma\, \ln t}{1-\beta} - \frac{\beta\gamma\, \ln R(t)}{1-\beta^2} \\
+\frac{2\sqrt{K}}{\sqrt{1-\beta^2}} {\rm atan}\frac{\left(1+\beta\right)t+\gamma}{\sqrt{1-\beta^2}\sqrt{K}} + C, \label{eq:ei} \end{multline}
with $K$ defined in Eq.~\eqref{eq:k} and the substitution variables explicitly given by
\begin{equation} t=u+\sqrt{u^2+\alpha^2} ,\end{equation}
and
\begin{equation} R(t) = 2 t T(u) = 2\left( u+\sqrt{u^2+\alpha^2} \right) T(u). \end{equation}

To illustrate the application of this result, consider an arbitrary vector ${\mathbf r}$ and two non-collinear unit vectors ${\mathbf e}_a$ and ${\mathbf e}_b$ (i.e., ${\mathbf e}_a\times {\mathbf e}_b \neq {\mathbf 0}$). 
We define the parameters:
\begin{align} 
\alpha &= \left| {\mathbf r} - {\mathbf e}_b \left( {\mathbf r}\cdot {\mathbf e}_b\right) \right|, \\
\beta &= - {\mathbf e}_a \cdot {\mathbf e}_b, \\
\gamma &= {\mathbf r}\cdot {\mathbf e}_a - \left( {\mathbf e}_a \cdot {\mathbf e}_b \right) \left( {\mathbf r}\cdot {\mathbf e}_b \right). 
\end{align}
These parameters yield
\begin{equation} 
K = \frac{\left| {\mathbf r}\cdot \left( {\mathbf e}_a \times {\mathbf e}_b \right) \right|^2}{\left| {\mathbf e}_a \times {\mathbf e}_b \right|^2}, 
\end{equation}
which naturally satisfies the conditions $K > 0$ and $1 - \beta^2 > 0$. 
Geometrically, $\sqrt{K}$ represents the perpendicular distance from the point ${\mathbf r}$ to the plane spanned by ${\mathbf e}_a$ and ${\mathbf e}_b$.

%%%%%%%%%%%%%%%%%%%%%%%%%%%%%%%%%%%%%%%%%%%%%%%%%%%%%%%%%%%%%%%%%%%%%%%%%%%%%%%%
\section{Appendix B: Electrostatic potentials of uniformly charged geometries}
\renewcommand{\theequation}{B\arabic{equation}}  % ← key line
\setcounter{equation}{0}                          % restart numbering
\begin{figure}[!htb]\centerline{\includegraphics[width=7.5cm]{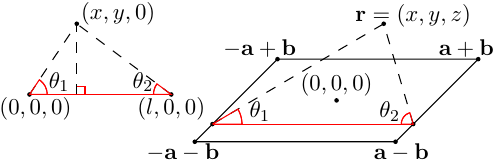} }          %---------------------  Figure -------------------------------------
\caption{Electrostatic potentials of (left) a uniformly charged straight line segment lying on the $x$-axis, and (right) a uniformly charged parallelogram with vertices at $-{\mathbf a}-{\mathbf b}$, ${\mathbf a}-{\mathbf b}$, ${\mathbf a}+{\mathbf b}$, and $-{\mathbf a}+{\mathbf b}$. 
For the line segment, the potential at any field point is determined by the angles $\theta_1$ and $\theta_2$ subtended by the endpoints of the segment [see Eq.~\eqref{eq:logtheta}].
\label{fig:chgline}}\end{figure}

To derive the electrostatic potential of a uniformly charged line segment, we first align the segment of length $l$ with the $x$-axis, extending from the origin to $(l,0,0)$ (see the left panel of Fig.~\ref{fig:chgline}). 
Consider a field point located at $\mathbf{r}=(x,y,0)$ in the same plane. Denoting the electrostatic potential by $\Phi$, we have
\begin{equation}  \Phi = \frac{Q}{l} \int_0^l \frac{du}{\sqrt{(x-u)^2 + y^2}} = \frac{Q}{l}\int_{-x}^{l-x} \frac{d\tau}{\sqrt{\tau^2+y^2}}, \end{equation}
where $Q/l$ is the charge density and the substitution $\tau = u-x$ has been applied. 
Evaluating this integral via the Euler substitution $\sqrt{\tau^2+y^2} + \tau = t$, which implies 
\begin{equation} \tau = \frac{1}{2}\left(t - \frac{y^2}{t}\right); \quad\quad \frac{d\tau}{\sqrt{\tau^2+y^2}} = \frac{dt}{t}, \end{equation}
yields the closed-form expression
\begin{equation}  \Phi = \frac{Q}{l} \ln \left[ \frac{\sqrt{(l-x)^2 + y^2} + l - x}{\sqrt{x^2+y^2} - x} \right]. \end{equation}
This result can be recast into a more intuitive geometric form by introducing the angles $\theta_1$ and $\theta_2$ between the segment and the vectors connecting the field point to the two endpoints:
\begin{equation} \Phi = \frac{Q}{l} \left[ \ln\left(\frac{1+\cos\theta_1}{\sin\theta_1}\right) + \ln\left(\frac{1+\cos\theta_2}{\sin\theta_2}\right) \right]. \label{eq:logtheta} \end{equation}
Each logarithmic term corresponds to the contribution from the right triangle formed by projecting the field point onto the line. 
The sign of each term depends on the field point's position relative to the segment; contributions are positive when the corresponding angle is acute [$\theta_1, \theta_2 \in (0, \pi/2)$] and negative when obtuse. 
Using standard trigonometric identities, each factor can be equivalently expressed as
\begin{equation} 
\frac{1+\cos\theta_1}{\sin\theta_1} = \frac{\sin\theta_1}{1-\cos\theta_1} = \cot\frac{\theta_1}{2}, \quad \mbox{etc.},
\end{equation}

For a general three-dimensional configuration, consider a uniformly charged line segment with length $l=|{\mathbf r}_1 - {\mathbf r}_2|$, defined by arbitrary endpoints ${\mathbf r}_1$ and ${\mathbf r}_2$.
Assuming a unit charge density ($Q/l = 1$), the electrostatic potential at an arbitrary field point ${\mathbf r} = (x,y,z)$ is given by the line integral
\begin{equation} \psi({\mathbf r}_1,{\mathbf r}_2,{\mathbf r}) = \int_0^1 \frac{dt\,\left|{\mathbf r}_2-{\mathbf r}_1\right| }{\left| {\mathbf r}_1 + t\left( {\mathbf r}_2-{\mathbf r}_1\right) - {\mathbf r} \right|}. \end{equation} 
To evaluate this geometrically, we introduce the unit vector along the segment $\hat{{\mathbf r}}_{12} = ({\mathbf r}_1 - {\mathbf r}_2)/l$ (pointing from ${\mathbf r}_2$ to ${\mathbf r}_1$) and its opposite $\hat{{\mathbf r}}_{21} = -\hat{{\mathbf r}}_{12}$.
The geometry is characterized by the angles $\theta_1$ and $\theta_2$; $\theta_1$ is the angle between the two vectors: ${\mathbf r}-{\mathbf r}_1$ and ${\mathbf r}_2-{\mathbf r}_1$ (i.e., the interior angle at vertex ${\mathbf r}_1$), and $\theta_2$ is defined analogously at ${\mathbf r}_2$.
The two angles satisfy
\begin{equation} 
\frac{\sin\theta_1}{\sin\theta_2} = \frac{|{\mathbf r}_2 - {\mathbf r}|}{|{\mathbf r}_1 - {\mathbf r}|},
\end{equation}
and
\begin{equation} 
\cos\theta_1 = \frac{({\mathbf r} - {\mathbf r}_1) \cdot \hat{{\mathbf r}}_{21}}{|{\mathbf r} - {\mathbf r}_1|}, \quad
\cos\theta_2   = \frac{({\mathbf r} - {\mathbf r}_2) \cdot \hat{{\mathbf r}}_{12}}{|{\mathbf r} - {\mathbf r}_2|}.
\end{equation}
Substituting these geometric relations into the closed-form expression (Eq.~\eqref{eq:logtheta}) yields the compact vector expressions
\begin{equation}  \psi({\mathbf r}_1,{\mathbf r}_2,{\mathbf r}) = \ln \left[ \frac{|{\mathbf r} - {\mathbf r}_1| + ({\mathbf r} - {\mathbf r}_1) \cdot \hat{{\mathbf r}}_{21}}{|{\mathbf r} - {\mathbf r}_2| + ({\mathbf r} - {\mathbf r}_2) \cdot \hat{{\mathbf r}}_{21}} \right], \label{eq:log1}
\end{equation}
or, equivalently,
\begin{equation} \psi({\mathbf r}_1,{\mathbf r}_2,{\mathbf r}) = \ln \left[ \frac{|{\mathbf r} - {\mathbf r}_2| + ({\mathbf r} - {\mathbf r}_2) \cdot \hat{{\mathbf r}}_{12}}{|{\mathbf r} - {\mathbf r}_1| + ({\mathbf r} - {\mathbf r}_1) \cdot \hat{{\mathbf r}}_{12}} \right]. \label{eq:log2}
\end{equation}
These formulas are universally valid for any interaction governed by an inverse-distance law; by replacing the total charge with the total mass, they equally describe the gravitational potential of a uniform thin rod.

We now proceed to derive the electrostatic potential of a uniformly charged parallelogram with total charge $Q$.
The surface is spanned by two basis vectors ${\mathbf a}$ and ${\mathbf b}$, with vertices located at $-{\mathbf a}-{\mathbf b}$, ${\mathbf a}-{\mathbf b}$, ${\mathbf a}+{\mathbf b}$, and $-{\mathbf a} + {\mathbf b}$ (see the right panel of Fig.~\ref{fig:chgline}).
The parallelogram is centered at the origin and has a total surface area $S=4\left| {\mathbf a}\times{\mathbf b}\right|$.
For a unit surface charge density $Q/S=1$, the potential at an arbitrary field point ${\mathbf r}=(x,y,z)$ can be expressed as a double integral
\begin{equation} \phi({\mathbf a},{\mathbf b},{\mathbf r}) = \int_{-1}^1 d\tau \int_{-1}^1 dt\frac{\left|{\mathbf a}\times{\mathbf b}\right|}{\left|t{\mathbf a}+\tau{\mathbf b}-{\mathbf r}\right|}, \label{eq:phidi} \end{equation}
where $t$ and $\tau$ are dimensionless parameters that sweep the parallelogram.
From this definition, it is straightforward to verify that $\phi({\mathbf a},{\mathbf b},{\mathbf r})$ is even in each of the vectors ${\mathbf a}$, ${\mathbf b}$ and ${\mathbf r}$, and symmetric under the exchange of ${\mathbf a}$ and ${\mathbf b}$; that is,
\begin{equation} \phi({\mathbf a},{\mathbf b},{\mathbf r}) = \phi(-{\mathbf a},{\mathbf b},{\mathbf r}) = \phi({\mathbf b},{\mathbf a},-{\mathbf r}), \quad\text{etc.}.  \end{equation}
The double integral can be evaluated by recognizing that the parallelogram may be decomposed into a continuum of parallel line segments aligned with ${\mathbf a}$.
Integration over $t$ first recovers the potential of a line segment, as derived in Eqs.~\eqref{eq:log1} and~\eqref{eq:log2}.
Consequently, the potential reduces to one-dimensional integral of logarithmic functions,
%\begin{equation} \phi({\mathbf a},{\mathbf b},{\mathbf r}) =\frac{\left|{\mathbf a}\times{\mathbf b}\right|}{2\left|{\mathbf a}\right|}\int_{-1}^1d\tau\, \psi(-{\mathbf a}+\tau{\mathbf b},{\mathbf a}+\tau{\mathbf b},{\mathbf r}),  \end{equation}
%where $\psi(-{\mathbf a}+\tau{\mathbf b},{\mathbf a}+\tau{\mathbf b},{\mathbf r})$ represents the potential of a line segment extending from $-{\mathbf a}+\tau{\mathbf b}$ to ${\mathbf a}+\tau{\mathbf b}$ (cf. Fig.~\ref{fig:chgline})
%\begin{equation} \psi(-{\mathbf a}+\tau{\mathbf b},{\mathbf a}+\tau{\mathbf b},{\mathbf r}) \end{equation}
\begin{multline} \phi({\mathbf a},{\mathbf b},{\mathbf r}) =\frac{\left|{\mathbf a}\times{\mathbf b}\right|}{2\left|{\mathbf a}\right|}\int_{-1}^1d\tau\, \psi(-{\mathbf a}+\tau{\mathbf b},{\mathbf a}+\tau{\mathbf b},{\mathbf r}) \\
= \frac{\left|{\mathbf a}\times{\mathbf b}\right|}{2\left|{\mathbf a}\right|}\int_{-1}^1d\tau \ln\frac{\left|{\mathbf r}+{\mathbf a}-\tau {\mathbf b}\right| + \left({\mathbf r}+{\mathbf a}-\tau {\mathbf b}\right)\cdot{\mathbf e}_a}
{\left|{\mathbf r}-{\mathbf a}-\tau {\mathbf b}\right| + \left({\mathbf r}-{\mathbf a}-\tau {\mathbf b}\right)\cdot{\mathbf e}_a},  \end{multline}
where ${\mathbf e}_a = {\mathbf a}/\left|{\mathbf a}\right|$ is the unit vector along ${\mathbf a}$, and the logarithmic term represents the potential of a line segment extending from $-{\mathbf a}+\tau{\mathbf b}$ to ${\mathbf a}+\tau{\mathbf b}$ (cf. Fig.~\ref{fig:chgline}).
For a general charge density, the result is recovered by multiplying the above expression by $Q/S$.

To evaluate the remaining integral involving logarithmic terms, we consider the general indefinite form
\begin{equation} I = \int d\tau\, \ln\left[ \left|{\mathbf c}-\tau {\mathbf b}\right| + \left({\mathbf c}-\tau {\mathbf b}\right)\cdot{\mathbf e}_a  \right],  \end{equation}
where ${\mathbf c}={\mathbf r}\pm {\mathbf a}$.
Applying the linear substitution
\begin{equation} b\tau = u + {\mathbf c}\cdot{\mathbf e}_b; \quad\text{with}\quad b=\left|{\mathbf b}\right|,\quad {\mathbf e}_b= {\mathbf b}/b, \end{equation}
transforms the integral into
\[b^{-1} \int du\, \ln \left( \sqrt{u^2+\alpha^2} + \beta u + \gamma \right),  \]
with the parameters defined as
\begin{equation} \alpha = \left| {\mathbf c} - {\mathbf e}_b \left( {\mathbf c}\cdot {\mathbf e}_b\right) \right|; \quad \beta = - {\mathbf e}_a \cdot {\mathbf e}_b, \end{equation}
\begin{equation} \gamma = {\mathbf c} \cdot {\mathbf e}_a - \left( {\mathbf c}\cdot {\mathbf e}_b\right) \left( {\mathbf e}_a \cdot {\mathbf e}_b \right)\equiv  \left({\mathbf c}\times{\mathbf e}_b\right)\cdot \left({\mathbf e}_a \times {\mathbf e}_b \right). \end{equation}
This integral admits a closed-form antiderivative that yields a combination of arctangent and logarithmic terms, as derived in Eq.~\eqref{eq:alog}.
Transforming back to the original variables, the potential $\phi$ as a function of the three vectors is expressed as:
\begin{equation} \phi({\mathbf a},{\mathbf b},{\mathbf r}) = \phi_{\rm log}({\mathbf a},{\mathbf b},{\mathbf r}) + \phi_{\rm atan}({\mathbf a},{\mathbf b},{\mathbf r}), \label{eq:phi1} \end{equation}
where the logarithmic and arctangent contributions are given, respectively, by
%\begin{multline} \phi_{\rm log}({\mathbf a},{\mathbf b},{\mathbf r}) = \sum_{t,\tau=\pm 1} \tau\left(av + t r_1\right) \ln s_2(t,\tau) \\ + \sum_{t,\tau=\pm 1} t\left(bv - \tau r_2\right) \ln s_1(t,\tau), \end{multline}
%\begin{equation} \phi_{\rm log}({\mathbf a},{\mathbf b},{\mathbf r}) = \sum_{t,\tau=\pm 1} \tau\left(av + t r_1\right) \ln s_2(t,\tau) + \sum_{t,\tau=\pm 1} t\left(bv - \tau r_2\right) \ln s_1(t,\tau), \end{equation}
and 
\begin{equation} \phi_{\rm atan}({\mathbf a},{\mathbf b},{\mathbf r})= 2r_0 \sum_{t,\tau=\pm 1} t \tau\, {\rm atan}\frac{s_3(t,\tau)}{v r_0}. \label{eq:phia1} \end{equation}
Eqs.~\eqref{eq:phi1} through~\eqref{eq:phia1} are identical to Eqs.~\eqref{eq:phi} through~\eqref{eq:phia}.
The auxiliary scalar parameters, vectors, and functions are defined as follows.
The magnitudes and unit vectors of ${\mathbf a}$ and ${\mathbf b}$ are given by
\begin{equation} a = \left| {\mathbf a} \right|; \quad b = \left| {\mathbf b} \right|; \quad {\mathbf e}_a = {\mathbf a}/a; \quad {\mathbf e}_b = {\mathbf b}/b. \label{eq:ab1}  \end{equation}
The orthogonal components, their cross product, and the scalar magnitude are defined as:
\begin{align} 
{\mathbf v}_1 &= {\mathbf e}_a - {\mathbf e}_b \left( {\mathbf e}_a\cdot {\mathbf e}_b \right) \equiv {\mathbf e}_b\times \left( {\mathbf e}_a \times {\mathbf e}_b \right),  \\ 
{\mathbf v}_2 &= {\mathbf e}_b - {\mathbf e}_a \left( {\mathbf e}_a\cdot {\mathbf e}_b \right) \equiv {\mathbf e}_a\times \left( {\mathbf e}_b \times {\mathbf e}_a \right), \\ 
{\mathbf v}_0 &= {\mathbf e}_a \times {\mathbf e}_b; \quad v = \left| {\mathbf v}_0 \right| = \left| {\mathbf v}_1 \right| = \left| {\mathbf v}_2 \right|.  
\end{align}
%\begin{equation} {\mathbf v}_1 = {\mathbf e}_a - {\mathbf e}_b \left( {\mathbf e}_a\cdot {\mathbf e}_b \right) \equiv {\mathbf e}_b\times \left( {\mathbf e}_a \times {\mathbf e}_b \right), \end{equation}
%\begin{equation} {\mathbf v}_2 = {\mathbf e}_b - {\mathbf e}_a \left( {\mathbf e}_a\cdot {\mathbf e}_b \right) \equiv {\mathbf e}_a\times \left( {\mathbf e}_b \times {\mathbf e}_a \right), \end{equation}
%\begin{equation} {\mathbf v}_0 = {\mathbf e}_a \times {\mathbf e}_b; \quad v = \left| {\mathbf v}_0 \right| = \left| {\mathbf v}_1 \right| = \left| {\mathbf v}_2 \right|.  \end{equation}
The projections of ${\mathbf r}$ onto the directions of ${\mathbf v}_0$, ${\mathbf v}_1$ and ${\mathbf v}_2$ are given by
\begin{equation} r_0 = {\mathbf r}\cdot \frac{{\mathbf v}_0}{v};\quad r_1 = {\mathbf r}\cdot \frac{{\mathbf v}_1}{v}; \quad r_2 = {\mathbf r}\cdot \frac{{\mathbf v}_2}{v},  \end{equation}
The relative position vectors connecting the vertices to the field points and their magnitude are defined as
\begin{equation} {\mathbf r}_s(t,\tau) = {\mathbf r} + t {\mathbf a} - \tau {\mathbf b};\quad r_s(t,\tau) = \left| {\mathbf r}_s(t,\tau) \right|,  \end{equation}
where $t,\tau \in \{-1, 1\}$.
Finally, the auxiliary scalar functions are given by
\begin{align}
& s_1(t,\tau) = r_s(t,\tau) + {\mathbf e}_a \cdot {\mathbf r}_s(t,\tau), \\
& s_2(t,\tau) = r_s(t,\tau) - {\mathbf e}_b \cdot {\mathbf r}_s(t,\tau), \\
& s_3(t,\tau) = r_s(t,\tau)\left(1-{\mathbf e}_a\cdot{\mathbf e}_b\right) + {\mathbf r}_s(t,\tau)\cdot\left( {\mathbf e}_a - {\mathbf e}_b \right). \label{eq:s31}
\end{align}
%\begin{equation} s_1(t,\tau) = r_s(t,\tau) + {\mathbf e}_a \cdot {\mathbf r}_s(t,\tau),\end{equation} 
%\begin{equation} s_2(t,\tau) = r_s(t,\tau) - {\mathbf e}_b \cdot {\mathbf r}_s(t,\tau), \end{equation}
%and
%\begin{equation} s_3(t,\tau) = r_s(t,\tau)\left(1-{\mathbf e}_a\cdot{\mathbf e}_b\right) + {\mathbf r}_s(t,\tau)\cdot\left( {\mathbf e}_a - {\mathbf e}_b \right). \label{eq:s31} \end{equation}
The set of equations from~\eqref{eq:ab1} to~\eqref{eq:s31} are the same as that from~\eqref{eq:ab} to~\eqref{eq:s3}.
Certainly, when ${\mathbf r}$ lies in the plane spanned by ${\mathbf e}_a$ and ${\mathbf e}_b$, we have $r_0 = 0 $ and hence $\phi_{\rm atan} = 0$.
Thus, $\phi_{\rm atan}$ serves as a measure of the deviation of ${\mathbf r}$ from this plane, being nonzero only when ${\mathbf r}$ has a nonzero perpendicular component.
Numerical tests confirm that for arbitrary vectors ${\mathbf a}$, ${\mathbf b}$, and ${\mathbf r}$, the inequality $\phi_{\rm atan}({\mathbf a},{\mathbf b},{\mathbf r}) \leqslant 0 $ always holds.

The symmetry properties of $\phi$ are satisfied independently by both $\phi_{\rm log}$ and $\phi_{\rm atan}$:
\begin{equation} \phi_{\rm log}({\mathbf a},{\mathbf b},{\mathbf r}) = \phi_{\rm log}(\pm {\mathbf a}, \pm {\mathbf b},{\mathbf r}) = \phi_{\rm log}({\mathbf b},{\mathbf a},{\mathbf r}),  \end{equation}
and 
\begin{equation} \phi_{\rm atan}({\mathbf a},{\mathbf b},{\mathbf r}) = \phi_{\rm atan}(\pm {\mathbf a}, \pm {\mathbf b},{\mathbf r}) = \phi_{\rm atan}({\mathbf b},{\mathbf a},{\mathbf r}).  \end{equation}
Nevertheless, direct verification of these identities from the explicit analytical expressions is non-trivial.

For the special case of a rectangular plate with the field point located on its central axis, we set ${\mathbf a}=(a,0,0)$, ${\mathbf b}=(0,b,0)$, and ${\mathbf r}=(0,0,z)$. 
Defining $R=\sqrt{a^2+b^2+z^2}$, the expression simplifies to
\begin{equation}\phi = 2a\log\frac{R+b}{R-b} + 2b\log\frac{R+a}{R-a} - 4z\,{\rm atan}\frac{ab}{zR}, \end{equation}
which has been obtained by employing the following arctangent identity:
\begin{multline} 2 {\rm atan}\frac{ab}{zR} + {\rm atan}\frac{R+a+b}{z} + {\rm atan}\frac{R-a-b}{z} \\ = {\rm atan}\frac{R-a+b}{z} + {\rm atan}\frac{R+a-b}{z}. \label{eq:iarc} \end{multline}
Furthermore, for a square plate ($a=b$), the potential reduces to $\phi({\mathbf a},{\mathbf b},{\mathbf r}) = 4a f(z/a)$, where the dimensionless function is given by
\begin{equation} f(x) = \log\frac{\sqrt{x^2+2}+1}{\sqrt{x^2+2}-1} - x \, {\rm atan}\frac{1}{x\sqrt{x^2+2}}. \end{equation}
This analytical result has been utilized to compare crystals of different macroscopic shapes, as well as crystals of the same shape but with different modes of replication of the unit cell.
These comparisons yield consistent results regarding the physical interpretation of the boundary term\cite{Zhao_Hu2026,He_Hu2026}.
%%%%%%%%%%%%%%%%%%%%%%%%%%%%%%%%%%%%%%%%%%%%%%%%%%%%%%%%%%%%%%%%%%%%%%%%%%%%%%%%
\section*{Data and Software Availability}
Both the data and the fortran code used to generate the data are openly available in github\cite{notedata}: http://github.com/zhonghanhu1981/mdcode.

%%%%%%%%%%%%%%%%%%%%%%%%%%%%%%%%%%%%%%%%%%%%%%%%%%%%%%%%%%%%%%%%%%%%%%%%%%%%%%%%%%%%%%%%%%%%%%%%%%%%%%%%%%%%%%%%%%%%%%%%%%%%%%%%%%%%%%%%%%%%%%%%%%%%%%%%%%%%%%%%
%\bibliography{reftric}

\end{document}